# Exploitation of Transparent Conductive Oxides in the Implementation of a Window-Integrated Wireless Sensor Node

Kaarle Jaakkola and Kirsi Tappura

*Abstract*—Exploitation of transparent conductive oxides (TCO) to implement an energy-autonomous sensor node for a wireless sensor network (WSN) is studied and a practical solution presented. In the practical implementations, flexible and rigid substrates that is polyimide and glass, are coated with TCO, namely aluminum doped zinc oxide (AZO). AZO-coated flexible substrates are used to form thermoelectric generators (TEG) that produce electricity for the sensor electronics of the node from thermal gradients on a window. As the second solution to utilize AZO, its conductive properties are exploited to implement transparent antennas for the sensor node. Antennas for a UHF RFID transponder and the Bluetooth radio of the node are implemented. A prototype of a flexible transparent TEG, with the area of 67 cm$^2$ when folded, was measured to produce power of 1.6 µW with a temperature difference of 43 K. A radiation efficiency of -9.1 dB was measured for the transparent RFID antenna prototype with the center frequency of 900 MHz. Radiation efficiencies between -3.8 dB and -0.4 dB, depending on the substrate, were obtained for the 2.45 GHz Bluetooth antenna.

## I. Introduction

IMPLEMENTATION of internet of things (IoT) sets requirements as well as gives possibilities for new concepts of wireless sensing. Small and low-cost electronics components, such as energy harvesting circuits and systems-on-chip (SoC) combining microcontrollers with integrated radio transceivers, are increasingly available, which helps to implement various types of sensor nodes to form wireless sensor networks (WSN) [1]. However, one of the greatest challenges in the way of implementing wireless sensor nodes to new environments is still the power generation; most of the current small sensor nodes are operated with a battery. An ideal maintenance-free sensor node, instead, would work without the need of battery change. Therefore, different concepts of energy harvesting have been studied in order to develop an energy autonomous wireless sensor node [2]. In the following, a solution that generates its electricity from temperature gradients on a window is presented. The concept is based on utilizing flexible foils on which a thin layer of transparent conductive oxide (TCO) is implemented as thermoelectric generators (TEGs). Powering WSN nodes by TEGs has been studied formerly [3], [4], but the novelty of the proposed solution lies in utilizing light, flexible and transparent foils that have also other uses in the sensor concept. In this study, additionally to using the transparent foils as TEGs, their conductive properties are exploited by implementing antennas on them, resulting as transparent antennas to be integrated on a window. Different types of surfaces can be coated with TCO. Therefore, additionally to flexible foils, glass is also studied here as the substrate of a transparent antenna. Implementing antennas on a window instead of the printed circuit board (PCB) reduces the size of the electronics of the sensor node and using TCO makes them genuinely transparent, in contrast to e.g. wire mesh solutions that provide an option to implement partly transparent antennas [5].

The third way of using the foils coated with TCO in the WSN concept is using their thermoelectric properties for touch sensing. This paper, however, concentrates on the two parts first mentioned: the thermoelectric generators and the transparent antennas. The touch sensor utilizing thermoelectric elements on TCO is based on the same phenomenon as their use as TEGs and is presented in detail in [6]. Adaptation of the electronics components of the node to operate with the devices implemented on TCO is also addressed.

Connecting the TEG elements processed on TCO with each other and to the electronics requires wiring with better electrical conductivity than what can be achieved with TCO materials. This wiring is implemented by screen-printing with silver nanoparticle ink.

The practical implementation of low-energy sensor electronics is used here as a part of the system and as a reference for power consumption. However, the optimization of the power consumption of the electronics part is not in the scope of this paper.

Manuscript submitted Jan 21, 2018, revised May 11, 2018. This work has been part of the TransFlexTeg project, which has received funding from the European Union's Horizon 2020 research and innovation program 2014-2018 under grant agreement No 645241. The partial funding from VTT Technical Research Centre of Finland Ltd is also gratefully acknowledged.

The authors are with the VTT Technical Research Centre of Finland Ltd., FI-02044 Espoo, Finland (e-mail: firstname.surname@vtt.fi).

## II. MATERIALS AND METHODS

### A. Topology of the sensor node

The overall sensor concept can be divided into two main parts: the components implemented on TCO and the electronics module. The block diagram of the system is shown in Fig. 1. On the left hand side, surrounded by a solid line, are the components integrated on TCO: the thermoelectric generators or TEGs, the two antennas that is the RFID tag and the Bluetooth antenna, and the thermoelectric sensors. On the right hand side, surrounded by a dashed line, are the components of the electronics module.

The concept comprises two types of transparent antennas: 2.45 GHz Bluetooth antenna for data transfer from the sensor node to an external gateway device and an RFID transponder antenna operating between 865 and 928 MHz for global operation according to ISO 18 000 - 6C (EPC Gen2) standard. The passive RFID tag integrated into the window provides the window and the sensor node with an individual identification code, but optionally also an alternative way of exciting the sensors when there is no power available from the thermoelectric generators or from the energy storage. The backscatter modulation of the RFID protocol also provides an energy efficient option for wireless communication [7].

The heart of the electronics module is the Nordic Semiconductors nRF51422 system-on-chip with an integrated microcontroller and a Bluetooth radio transceiver unit [8]. As the power production of the TEGs, especially when installed on a window, varies over time, an energy harvesting circuit with a storage capacitor is assembled between the TEGs and the power consuming electronics. As the selected energy harvesting circuit, Analog Devices ADP5090, has a unipolar input and the polarity of TEGs assembled on a window can change depending on the time of the year and even within a day, there is a diode bridge consisting of four SD103 Schottky diodes to provide the energy harvesting circuit with a fixed polarity. In order to exploit the energy of the storage capacitor as efficiently as possible and to protect the components from overvoltage, a chopper regulator is assembled between the energy harvesting circuit and the power consuming electronics.

Three sensor units are connected to the nRF51422: HYT271 that measures the ambient temperature and relative humidity, GSS Cozir $CO_2$ sensor unit and the thermoelectric touch sensors integrated on TCO. As the voltage produced by the thermoelectric touch sensors is very low, only a few hundred microvolts, an instrumentation amplifier is needed between the sensors and the input channels of the nRF51422. As the touch sensor is a multichannel array, this amplifier unit also comprises a time-domain multiplexer (Mux). nRF51422 transmits the sensor data to the gateway device using Bluetooth 4.0 LE protocol. The gateway device connected to the Internet sends the measurement data further as http messages into a cloud service.

The average power consumption of the electronics is controlled by adjusting the duty cycle of the power hungry operations that is transmission and measurements. The idle power consumption of the CPU is 2 µW and the highest peak consumption during the 20 ms transmit burst is 60 mW. As an example of a use scenario, if temperature and humidity are measured and the data is transmitted every 15 minutes, the average power consumption of the electronics becomes 3.3 µW.

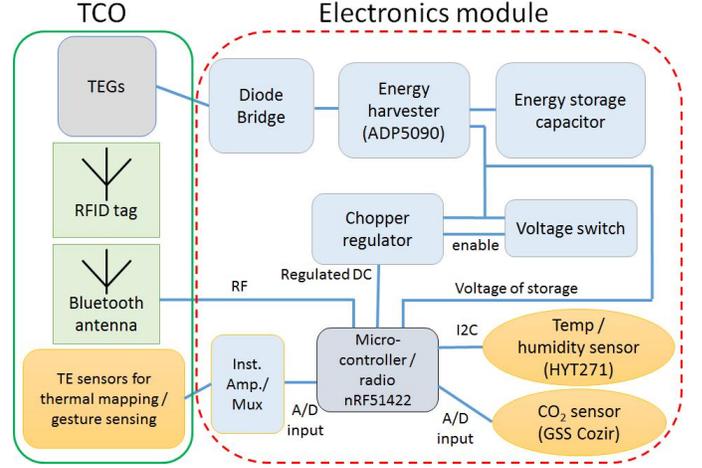

Fig. 1. Topology of the sensor node.

### B. Transparent conductive oxides

The figure of merit, *ZT,* for a thermoelectric material is defined [9]:

$$ZT = \frac{\sigma S^2 T}{\kappa}, \qquad (1)$$

where $\sigma$ is electrical conductivity, $S$ the Seebeck coefficient, $T$ absolute temperature and $\kappa$ thermal conductivity.

In terms of the critical parameters, the two applications of TCO studied in this paper, thermoelectric generators and antennas, have somewhat different requirements. For TEGs, the optimization of the material parameters is about finding the right combination or compromise between $\sigma$, $S$ and $\kappa$. For antennas, instead, the electrical conductivity $\sigma$ is the only critical parameter. Therefore, the TCOs need to be optimized primarily for their use as TEGs.

There are some alternatives for the TCO material, such as impurity-doped $SnO_2$, $In_2O_3$ and ZnO [10]. Tin doped indium oxide ($In_2O_3$:Sn or ITO) is one of the most commonly used TCOs due to its high electrical conductivity and high transparency, while aluminum doped zinc oxide (ZnO:Al or AZO) provides an environmentally friendly alternative that is more abundant and has lower cost. As AZO also has good electrical conductivity (for a TCO), high transparency (80–90% transmittance in the visible region [11]), reasonable thermoelectric properties [6], [12], [13], and very good RF properties considering the antenna solutions also at UHF [14], it is used in this work. Recently, some antennas have been implemented on ITO and AZO at the frequency of 2.45 GHz and above [15], [16]. However, lower frequencies such as 900 MHz studied in this paper are more challenging and require special measures to achieve an adequate radiation efficiency and to implement the electrical connection to the antenna.

The AZO coating of the prototypes was made by Picosun using an atomic layer deposition (ALD) process to enable sufficiently low process temperatures required by the flexible substrates [17]. Three materials were used as substrates for the AZO coating: regular polyimide or Kapton NH® film (yellow), Kapton CS film (colorless polyimide) and glass [18]. All the TEG and RFID transponder prototypes were made using Kapton NH film. Bluetooth antenna prototypes were made using Kapton CS film and glass. Regardless of the substrate, the AZO coating is ~ 400 nm thick, producing DC sheet resistance of about 40 to 55 Ω /□, slightly varying over the surface. The selected thickness of AZO is a result of a compromise between the thickness limitations of the ALD process and the requirements of the antenna and TEG applications for a sufficiently low sheet resistance.

### C. Thermoelectric generators.

The concept of harvesting energy from thermal gradients of a window was studied by measuring temperatures on the two inner glasses of a triple glazed window located in an office building in Espoo, Finland (60°11'8"N, 24°49'22"E), facing towards South-West. The air gap between these glasses is 12 mm. Fig. 2 a) shows the temperature difference between the glasses under winter conditions in January 2016, whereas Fig. 2 b) demonstrates the temperature differences available in summer (August 2015). The different curves in the figure stand for the different lateral positions of the pt-100 sensors on the glass. During the peak difference of winter, the temperature of the outer glass was -23.2 °C and during the peak of the opposite direction in summer it was +51.3 °C. These two cases of summer and winter also show how the polarity changes due to the change of the direction of the thermal gradient, justifying the need of the diode bridge between the TEGs and the unipolar energy harvester circuit. During summer time, the polarity changes also within a day. The temperature difference between the glasses is 23.8 K at its highest, but it can also be practically zero over rather long periods, based on which the energy storage needs to be sized.

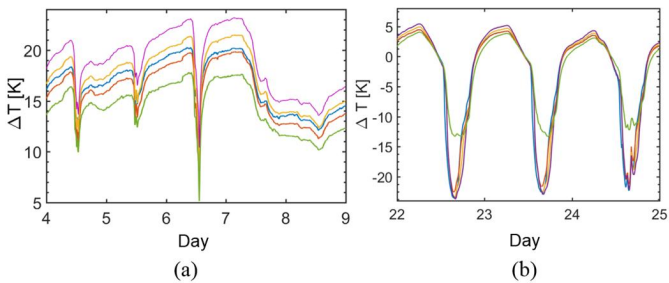

Fig. 2. Examples of temperature differences between the window glasses in (a) winter (January 2016) and (b) in summer (August 2015).

A thin-film TEG design with a novel folding scheme has been proposed previously for large-area, low energy density applications and the performance analyzed computationally [9]. The proposed design suits well for energy harvesting from the thermal gradients available between window glasses. In the folded TEG module, the heat flux and current flow are parallel to film surface but the temperature gradient perpendicular to the plane of the TE module, as described in [9]. The basic three-dimensional structure of the folded module designed to fit tightly in the 12 mm wide space between the glasses is shown in Fig. 3 (a). However, in order to avoid the possible risks of the folding process [19] and to enable easy control of thermal gradients, the present study concentrates on measuring the generated voltage and power of planar TEG foils but having the conductor and TEG patterns applicable to the folded structure (see Fig, 3 (b)).

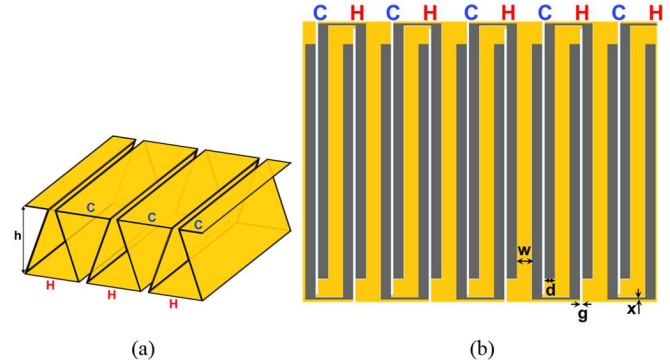

Fig. 3. (a) The principle of the proposed 3D structure of the folded TEG module. (b) The conductor and TEG patterns, for nine single-conduction-type TEGs connected electrically in series, applicable to the folded structure (yellow = AZO, grey = conductor lines, light blue lines = AZO removed, C = cold, H = hot). h = 12 mm, w = 8.05 mm, d = 5.1 mm, g = 1.4 mm, x = 1 mm.

For predicting the performance of the planar TEGs and for extracting the material parameters (see Section IIIA) of the fabricated thin film TEGs, a simulation model mimicking the planar measurement setup of Figs. 4 and 5 was built. The computational model is based on the finite element method (FEM) implemented in COMSOL Multiphysics [20]. In addition, analytical methods are used for calculating the selected characteristic parameters of the TE device. In the FEM model, heat transfer equations are coupled with the electrical phenomena for modelling the thermoelectric effect (Peltier-Seebeck-Thomson) [20]. The FEM model includes the coupled phenomena of heat transfer by conduction in the substrate, thermoelectric materials and conductor lines, electrical conduction and Joule heating in the TE material and conductor lines, and thermoelectric effect in the TE material and conductor lines. Temperature gradients are applied over the TEGs mimicking the temperature gradient of the folded structure by setting a constant heat source or constant temperature on the hot and cold lines as shown in Fig. 3 (b).

The TEG foil prototypes were fabricated by screen-printing the conductor pattern of Fig. 3 (b) with Inkron IPC-114 silver ink on the AZO-coated Kapton NH foils. AZO was removed from between the adjacent conductor lines at the same temperature, as shown in Fig. 3 (b), by grinding with fine sand paper.

In order to measure the voltage and power generated by the TEG foils, a test rig with aluminum fingers acting as heat conductors was built. Side view of the structure is shown in Fig. 4. In the structure, two aluminum plates (1) are thermally connected to aluminum fingers (2) that are each pressed against the foil under test (3) by a finger of synthetic rubber foam (4) on the opposite side of the foil. Five of the fingers are equipped with pt-100 temperature sensors (5) to measure the temperature difference. The structure is placed on a hot plate and is cooled from above e.g. by using ice. Every second pair of the foil is then at high temperature and the ones between them are at low temperature. In order to measure each thermopair also individually, there are also galvanic contacts formed by strips of copper tape on the upper part of the device. Fig. 5 shows the opened structure with the TEG foil under test between the halves of the device.

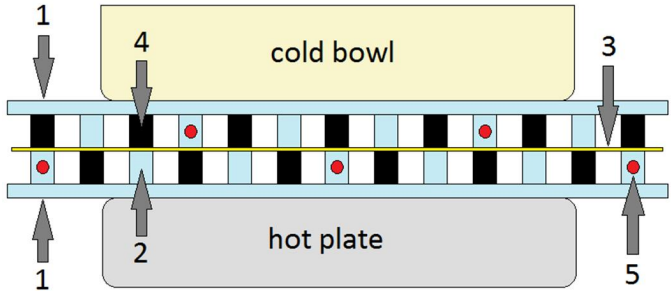

Fig. 4. Schematic side view of the thermoelectric test rig, showing the aluminum plates (1), aluminum fingers (2), foil under test (3), foam fingers (4) and temperature sensors (5).

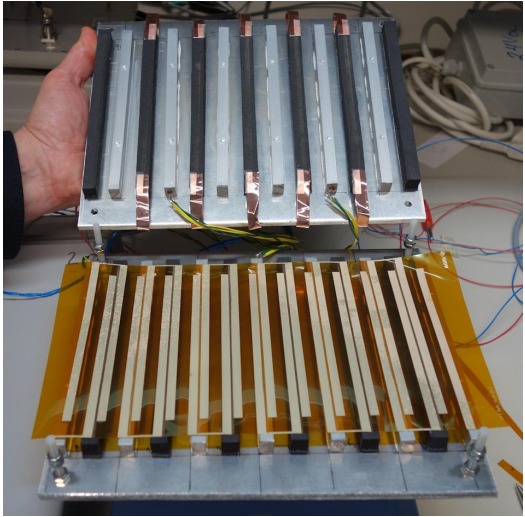

Fig. 5. Thermoelectric test rig opened with the TEG foil under test in place.

In the measurement of a TEG foil, the hot plate was heated up to 55 °C, while some ice was put on a bowl on the top of the test rig. Nine thermopairs of the twelve in total on the foil were connected to Agilent 34970A data acquisition unit via Agilent 34901A multiplexer board to measure the voltage and the electric power generated. The pt-100 temperature sensors were connected to the same data acquisition unit and the data was collected using a LabView software. The series resistance and thus also the electric power produced by the TEG foil was measured by loading the nine elements connected in series with changing resistance values.

### D. Antennas

The two antenna types implemented on TCO, the RFID transponder and the Bluetooth antenna, have somewhat different specifications. Additionally to operating at different frequencies, the required input impedance as well as the physical connection interface to the electronics is very different for the two. The very low conductivity of the TCO material sets its limitations to the possible antenna structures; in order to minimize the conductivity losses, electric current should be distributed as evenly as possible on the antenna conductor. In practice, this means simple structures with wide conductors and no meandering [21]. If the electronics require a large enough printed circuit board, it can be used as a ground plane or the other half of an asymmetric dipole antenna. Fig. 6 shows the antenna structures that meet these criteria and therefore can be considered as potential TCO antennas. Type a) is a simple two-branch dipole fed from its middle. Type b) is a dipole antenna with an integrated parallel loop that is typically used in label type RFID transponders [22]. Type c) is a straight dipole fed inductively with an external loop on its side. Type d) uses the printed circuit board shown on the right as one half of the asymmetric dipole, as the second one is implemented on TCO. Implementing a galvanic contact to TCO is challenging and therefore it is preferred to use capacitive coupling with antenna types a), b) and d).

The antennas were simulated with Ansys HFSS 15.0 electromagnetic simulation tool [23]. The material parameters used in the simulations are listed in Table I, including relative permittivity $\varepsilon_r$, dielectric loss tangent $tan(\delta)$ and conductivity $\sigma$.

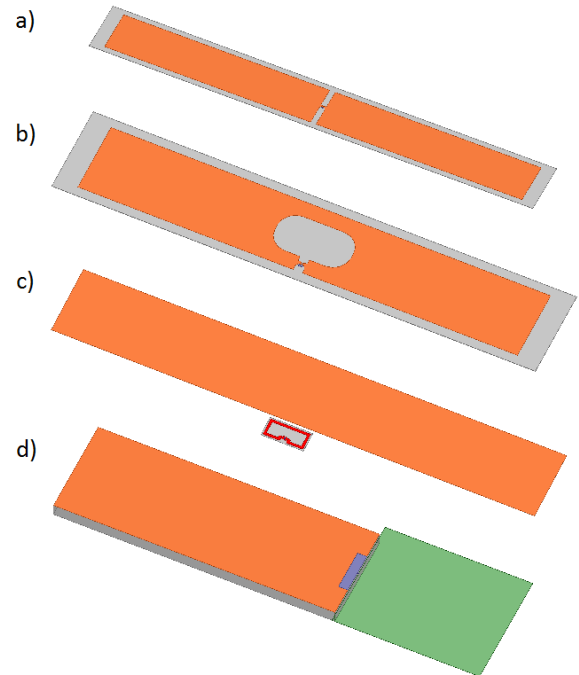

Fig. 6. Alternative antenna types implemented on TCO.

TABLE I
MATERIAL PARAMETERS USED IN THE ANTENNA SIMULATIONS

| thickness of AZO | | 400 nm | |
|---|---|---|---|
| Sheet resistance of AZO | | 53 Ω / □ | |
| other materials | $\varepsilon_r$ | $tan(\delta)$ | $\sigma$ |
| • Kapton NH/CS | 3.4 | 0.008 | |
| • glass | 5.5 | 0.005 | |
| • PET | 2.8 | 0.010 | |
| • FR4 epoxy | 4.4 | 0.020 | |
| • aluminum | | | $3.8 * 10^7$ S/m |

## III. RESULTS AND DISCUSSION

*A. Thermoelectric generators*

The measured resistance of the TE module consisting of nine TEGs connected in series (100 Ω) and the sheet resistance of AZO (53 Ω/□) were used as the starting points (known values) for the simulations. The measured series resistance of the module includes the resistance of AZO and silver ink lines as well as all the unknown contact resistances between the silver ink contact lines and AZO. First, the contact resistance was adjusted to produce the measured resistance of the module in the simulations. Then the Seebeck coefficient was varied to obtain the maximum measured output power. A good match was found with $S \approx -73$ μV/K that is in good accordance with the typical values of AZO reported in the literature [6], [12], [13]. The simulation results are shown in Fig. 7 where the output power is predicted for different temperature gradients as a function of load resistance. The parameters used in the simulations are listed in Table II. The thermal conductivity of AZO is also listed, although it has been shown that its influence on the device performance is negligible for such thin films [9].

TABLE II
PARAMETERS USED IN THE SIMULATIONS OF FIG. 7 AND 9.

| | AZO | Conductors | Kapton NH |
|---|---|---|---|
| **Measured (or literature [24]*, [18]**) values**: | | | |
| • Sheet resistance [Ω/□] | 53 | $7.5\times10^{-4}$ | |
| • Thickness [μm] | 0.38 | 50 | 25 |
| • Thermal conductivity, $\kappa$ [Wm$^{-1}$K$^{-1}$] | 3.5 | 238* | 0.12** |
| • Total resistance of 9 TEGs [Ω] | | 100 | |
| • Contact resistance [Ωcm$^2$] | | 30 | |
| **Calculated from the measured values:** | | | |
| • Electrical conductivity, $\sigma$ [S/m] | $5\times10^4$ | $2.67\times10^7$ | |
| **Obtained by fitting the simulations to the experimental data:** | | | |
| Seebeck coefficient, $S$ [μV/K] | -73 | 3.5 | |
| Power factor, PF [W/m/K$^2$] | $2.7\times10^{-4}$ | | |

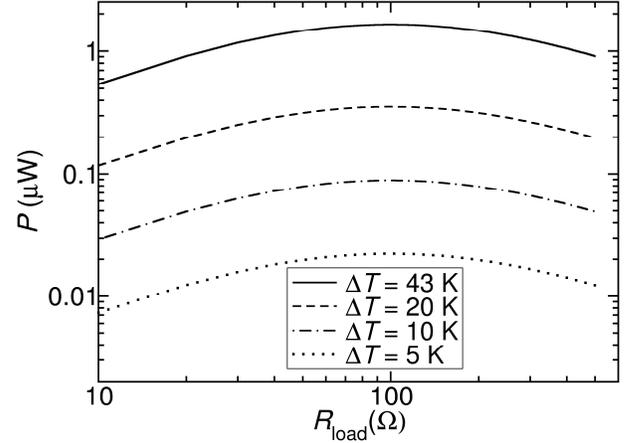

Fig. 7. The simulated DC power generated by the nine elements of a TEG foil in series for different temperature gradients as a function of load resistance.

Fig. 8 shows the temperatures and voltages as a function of time during the measurement with the test rig shown in Figs 4 and 5. The dashed lines represent the temperature differences I to IV; the temperature sensors shown in Fig. 4 are numbered by 1 to 5 from the left and the differences marked with roman numbers are between each neighboring pair (e.g. difference II is between sensors 2 and 3). The greatest temperature differences are in the middle of the rig (II and III) whereas differences I and IV on the sides are smaller. The solid lines represent voltages of each individual thermoelectric pair. The voltage level of these does not reflect the position of the pair, but seems to be "random", likely reflecting the quality of each pair.

After first measuring the open-circuit voltage, when the TEGs are loaded only by the 10 MΩ input resistance of the multimeter, resistive loads between 25 Ω and 2 kΩ were connected between the ends of the series connection of the nine TEG elements. The effect of connecting the loads can be seen in Fig. 8, where the presence of the load resistances, numbered from 1 to 7, is marked on the voltage curves in the middle of the horizontal part of the curve, where the resistor is connected. The resistance values from 1 to 7 are respectively 2 kΩ, 955 Ω, 512 Ω, 196 Ω, 100 Ω, 47 Ω and 25 Ω. Finally, after the resistance of 25 Ω (7.), the TEGs are measured once more in the open-circuit mode, which can be seen on the voltages raising back to their high values. The decrease of the open-circuit voltage between the start and the end of the measurement is taken into account in the calculations. Fig. 9 shows the results of the loading measurement that is the produced DC power as a function of load resistance $R_{load}$, compared to the corresponding simulated values. The highest power of 1.6 μW is obtained at $R_{load} \approx R_{TEG} \approx 100$ Ω, where $R_{TEG}$ is the measured resistance of the TE module consisting of nine TEGs, as expected.

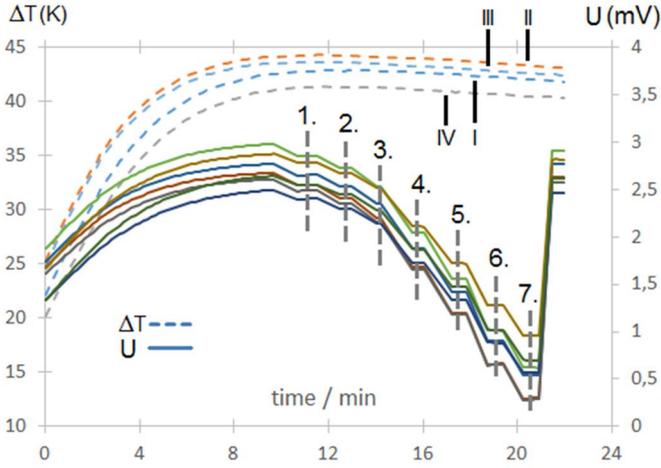

Fig. 8. Temperature difference and voltage of the TEG elements on the foil as a function of time over the measurement sequence, during which the load resistance is also varied.

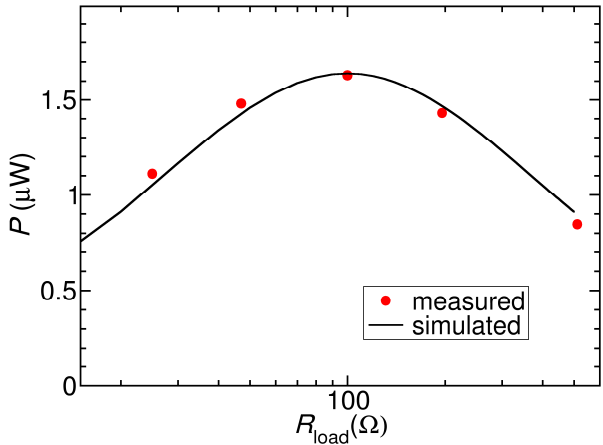

Fig. 9. DC power generated by the nine elements of a TEG foil in series as a function of load resistance; measured and simulated values.

When folded between glasses as shown in Fig. 3 (a), the area of the nine elements becomes 67 cm$^2$. Thus, a regular window glass of 0.5 m$^2$ can be equipped with 74 of these module elements, which produces about 118 µW with the 43 K temperature difference. However, using a more realistic long-time average temperature difference of 10 K (Fig. 2) and the simulation results of Fig. 7, the average power produced by such a window becomes 6.6 µW, which is still above the 3.3 µW power consumption of the use scenario given in Sec. II A. Although the produced power may seem small for the required area, the significant advantage of the proposed TEG design is that it minimizes the heat leakage through the module itself and, thus, maximizes the available temperature gradient under heatsink-limited conditions. Under similar conditions, the conventional TEGs can support only a fraction of the temperature gradient sustained by the proposed TEG, which in practice makes the former close to useless in the applications where efficient heat sinks cannot be used. This relates to the fact that the effective thermal conductivity of the folded TEG module is close to that of air, i.e. at least 30 - 150 times smaller than that of a conventional bulk TEG [25]. On the other hand, as the electrical current also flows in the plane of the thin AZO film, the electrical resistance of the proposed module gets high unless the aspect ratio of the legs is increased [9]. The Seebeck coefficient and electrical conductivity of AZO (Table II) are of the same order as those of many bulk TEGs, but inferior to those of the best materials.

Generally, one potential risk of the AZO coating is its durability during the folding or handling of the flexible substrates, which may cause cracks on the coating [19]. In order to evaluate primarily the electrical performance of the TEG prototypes without additional fabrication related risks, the TEG foils were measured in a planar form with a special test rig. However, based on the literature [19] as well as the preliminary tests performed by the authors, the proposed folding scheme seems feasible as far as the cracking sensitivity is taken into account in the fabrication process and the sharpest bends positioned on the metal lines. Another option is to perform the deposition of AZO in a later phase, i.e. on the folded substrate and, thus, to avoid the need to handle the flexible substrate with AZO on it. For ALD this is a valid option, as it produces conformal thin films regardless of the direction of the targeted surface.

Due to the limited availability of colorless Kapton CS, the TEG prototypes were made on yellow Kapton NH. This together with the wide silver conductors compromises the transparency of a window equipped with the TEG module. However, in these first prototypes the area of the silver conductor was not optimized and it is expected that, especially if the contact resistance between silver and AZO can be reduced, narrower conductors can be used in the future.

### B. RFID Transponder

In a typical RFID transponder a small bare-die microchip is attached directly to an antenna inlay, which is also the case with the prototype studied here. Consequently, there is no ground area of a PCB available to be used as a part of the antenna, which means that the antenna type d) of Fig. 6 is not an option for an RFID transponder.

In order to achieve conjugate impedance match with the microchip that has a capacitive input impedance, the input impedance of the antenna needs to be inductive. Therefore, antenna type b) of Fig. 6 is commonly used with commercial label type transponders [22]. However, the simulations showed that with the low conductivity of the AZO film, the parallel loop of any size does not produce inductive input impedance as it does with higher conductivities [21]. This leaves antenna type c) with an external loop made of high-conductivity material the only viable option.

Further optimization of the antenna type c) showed that the highest radiation efficiency combined with the right input impedance can be achieved with the structure shown with its dimensions (in mm) in Fig. 10. The dimensions of the inductive coupling loop made of 17 µm thick aluminum on a PET substrate are shown in Fig. 11.

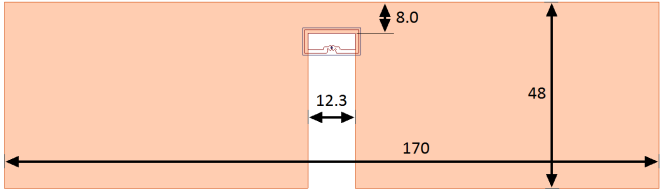

Fig. 10. Dimensions of the RFID transponder prototype in mm.

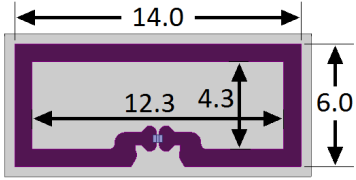

Fig. 11. Dimensions of the inductive coupling loop in mm.

The read range is a commonly used figure of merit for a passive UHF RFID transponder. The theoretical forward-link limited read range of the transponder can be calculated from the simulation results by:

$$R_{read} = \frac{c}{2\omega}\sqrt{\frac{P_{tx\,EIRP}D_{tag}\eta_{tag}\left(1-\left|\frac{Z_{tag}-Z_{IC}^*}{Z_{tag}+Z_{IC}}\right|^2\right)}{P_{IC\,sens}}}, \quad (2)$$

where $c$ is the speed of light, $\omega$ the angular frequency, $P_{tx\,EIRP}$ the equivalent isotropically radiated power of the reader device, $D_{tag}$ the directivity of the transponder antenna, $\eta_{tag}$ the radiation efficiency of the transponder antenna, $Z_{IC}$ the complex impedance of the microchip, $Z_{tag}$ the input impedance of the transponder antenna and $P_{IC\,sens}$ the read sensitivity of the microchip. '*' denotes complex conjugate. $P_{tx\,EIRP}$ = 3.28 W (2 W ERP), which is the maximum allowed radiated power of a UHF RFID reader as defined by ECC / ETSI [26]. The frequency-dependent impedance of the Monza R6 microchip $Z_{IC}$ is calculated, as specified by the manufacturer, by the parallel connection of chip resistance $R_p$ (1200 Ω), chip capacitance $C_p$ (1.23 pF) and mounting capacitance $C_{mount}$ (0.21 pF): $Z_{IC} = (R_p \parallel C_p \parallel C_{mount})$ [27]. The chip sensitivity $P_{IC\,sens}$ = -20 dBm [27]. The other parameters of (2) are obtained as simulation results as a function of frequency.

Four transponder prototypes were made using two sheets of AZO coated Kapton NH foils from different process batches. The prototypes are named 1A, 1B, 2A and 2B, with the number referring to the process batch. The prototypes were measured with Tagformance UHF RFID device using its own anechoic cabinet [28]. The transponder prototype inside the cabinet, supported by a piece of Styrofoam, is shown in Fig. 12. The evaluation is based on measuring the activation level of the transponder as a function of frequency in a fixed and known setup, which is normalized for each measurement series with a standard transponder, the frequency response of which is known [29]. As a result, the measurement gives the equivalent forward-link limited read range that is directly comparable with the values calculated by (2) from the simulation results. The simulated and measured read ranges of the transponders as a function of frequency are shown in Fig. 13.

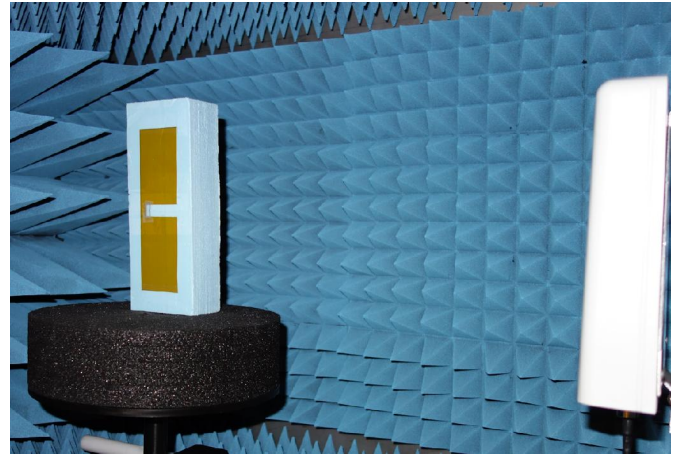

Fig. 12. Transponder prototype inside the measurement cabinet.

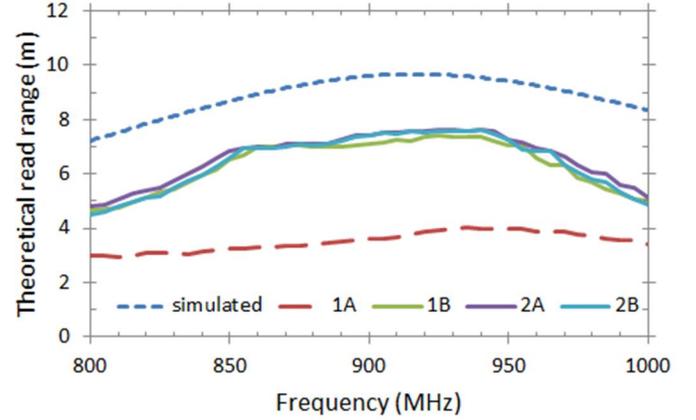

Fig. 13. Simulated and measured theoretical read range of the transponder prototypes as a function of frequency.

The graphs of Fig. 13 show that one of the prototypes, 1A, has a clearly lower read range than the rest three, which all have practically identical responses. Therefore, 1A can be excluded from the further analysis as a defective individual. The frequency band of the three is right for global operation, but their read range is shorter than that predicted by the simulation, namely 7.4 m vs. 9.6 m at 900 MHz. In terms of the power sensitivity of the transponder, the difference is 2.3 dB. The simulated radiation efficiency at 900 MHz is -6.8 dB (21 %). Consequently, if the difference in the sensitivity between the simulation and measurement results is explained by a difference in radiation efficiency, the measured radiation efficiency becomes -9.1 dB (12 %). A possible explanation for the difference is the transponder antenna being particularly sensitive to irregularities close to the coupling loop where the current density is at its highest [21]. The AZO coating on the edges of the antenna is likely to be somewhat irregular, compared to the smooth edges of the simulation model.

When compared to commercial label transponders made by etching of aluminum, the radiation efficiency of which is about

-0.5 dB (90 %) [22], the measured radiation efficiencies are quite low. However, the read range of 7.4 m is still adequate for many applications. By Eq. (2), the corresponding theoretical value for an antenna with -0.5 dB, the radiation efficiency is about 20 m [27].

*C. Bluetooth antenna*

The electronics module is built around an nRF51422 microchip and has also several other components that are all assembled on a 26 mm * 33 mm PCB. This PCB with its ground layers can be used as the second terminal of a dipole antenna, enabling the use of antenna type d) of Fig. 6 as the Bluetooth antenna. In order to combine the DC ground and the other terminal of the dipole antenna, BAL-NRF02D3 balun is connected between the nRF51422 microchip and the antenna. The required input impedance of the Bluetooth antenna is thus determined by the 50-Ohm output of the balun.

Two types of transparent antennas were studied: AZO on glass and AZO on Kapton CS. In the antenna prototypes, the output of the balun was connected to the antenna using a 9 mm * 2 mm copper strip. The size of the antenna was optimized and its expected performance evaluated by simulations.

Two antenna prototypes connected to the electronics module are shown in Fig. 14; the glass antenna on the left is attached to the PCB with a plastic clamp and the Kapton CS antenna is supported by a piece of Styrofoam and fixed to the PCB with a rubber band.

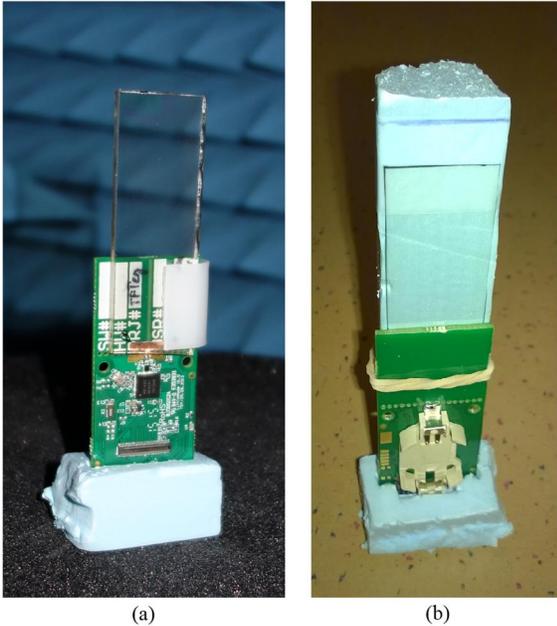

Fig. 14. Two Bluetooth antenna prototypes: Glass antenna (a) and flexible Kapton CS antenna (b).

In order to evaluate the antenna prototypes in terms of the radiation efficiency, they were measured in an anechoic cabinet. For the measurement, the Bluetooth module was programmed to continuously transmit carrier wave with 0 dBm power at the frequency of 2.45 GHz. The module was powered with a CR2032 Lithium battery to avoid any wires that would affect the antenna. The RF signal transmitted by the module was received with a Huber-Suhner 1324.19.0002 measurement antenna placed 0.45 m apart from the module in the cabinet. The received signal level was measured with Anritsu MS2830A spectrum analyzer. The radiation efficiency can then be calculated from the power budget of the measurement in decibel form:

$$\eta(dB) = P_{rx} - L_{rx} - G_{rx} - A_F - D_{tx} - L_b - L_Z - P_{tx}, \quad (3)$$

where $P_{rx}$ is the signal power measured by the spectrum analyzer, $L_{rx}$ the attenuation of the cable between the spectrum analyzer and the receiver antenna, $G_{rx}$ the gain of the receiver antenna, $A_F$ the free-space attenuation, $D_{tx}$ the directivity of the antenna prototype, $L_b$ the insertion loss of the balun, $L_Z$ the attenuation due to impedance mismatch between the balun and the antenna prototype and $P_{tx}$ the power transmitted by the Bluetooth module. Free-space attenuation can be calculated from the speed of light $c$, the frequency $f$ and the distance between the antennas $R$:

$$A_F(dB) = 20 \, log\left(\frac{c}{4\pi fR}\right). \quad (4)$$

$L_Z$ can be calculated from the output impedance of the balun $Z_b$ (50 Ω) and the complex input impedance of the antenna prototype $Z_A$:

$$L_Z = 1 - \left(\frac{Z_A - Z_b^*}{Z_A + Z_b}\right)^2. \quad (5)$$

The following values were used for (3): by measurement with a network analyzer $L_{rx}$ = -2.6 dB, $G_{rx}$ = 8.5 dBi for Huber-Suhner 1324.19.0002 [30], by (4) $A_F$ = -33.3 dB ($R$ = 0.45 m), $D_{tx}$ is determined by simulation, $L_b$ is -1.9 dB [31], $L_Z$ is calculated from the simulated input impedance of the antenna $Z_A$ using (5) and $P_{tx}$ is 0 dBm.

Four antenna prototypes were measured; two with a 3 mm glass substrate ("A" and "B"), one with a 1 mm glass substrate ("D") and one with a 50 μm thick Kapton CS substrate. The substrates are coated with AZO on the both sides, so the three glass prototypes were measured with the both sides touching the coupling strip, leading to seven measurement cases in total. Simulated and measured parameter values of the antenna prototypes are listed in Table III; simulated input impedance, simulated directivity in the direction of the measurement antenna in the test setup, simulated radiation efficiency and the measured radiation efficiency calculated using (3).

TABLE III
SIMULATED AND MEASURED ANTENNA PARAMETERS @ 2.45 GHZ.

| | simulated | | | | meas. |
|---|---|---|---|---|---|
| | $Z_A$ (Ω) | | $D_{tx}$ | $\eta$ | $\eta$ |
| sample, substrate - side | R | X | (dBi) | (dB) | (dB) |
| "A", Glass 3 mm - side 1 | 31 | -12 | 2.3 | -2.8 | -0.4 |
| "A", Glass 3 mm - side 2 | ,, | ,, | ,, | ,, | -1.6 |
| "B", Glass 3 mm - side 1 | ,, | ,, | ,, | ,, | -1.0 |
| "B", Glass 3 mm - side 2 | ,, | ,, | ,, | ,, | -2.0 |
| "D", Glass 1 mm - side 1 | 30 | -15 | 2.4 | -2.6 | -2.7 |
| "D", Glass 1 mm - side 2 | ,, | ,, | ,, | ,, | -3.6 |
| Kapton CS 50 μm | 29 | -17 | 2.5 | -2.2 | -3.8 |

The results of Table III show that the measurements actually give higher radiation efficiencies than the simulations for the antenna prototypes implemented on the thick glass substrate; for thin glass the simulation and measurement results are quite close to each other and for Kapton CS substrate the measured value is 1.6 dB lower than that given by the simulation. As AZO coating is known to be somewhat brittle [19], one may assume that the rigid glass as a stable substrate ensures a more homogeneous coating.

The achieved radiation efficiencies are comparable with or, as in the case of 3 mm glass substrate, better than the values of commercial chip antennas that are commonly used with Bluetooth modules. For such, -3 dB (50 %) is a typical value [32].

## IV. Conclusion

The use of Al-doped Zinc oxide (AZO) to form thermoelectric generators and antenna conductors for an energy-autonomous wireless sensor node was studied and demonstrated. The operation of the both was first simulated and then verified by measurements on prototypes. The concept of harvesting energy from temperature differences on a window was first studied by measuring the temperature differences that occur between actual glasses of a window in Espoo, Finland.

According to the measurements, the fabricated TEG prototype with nine elements produced power of 1.6 µW with a temperature difference of 43 K. With a more realistic long-term temperature difference of 10 K, simulations predict the power of 90 nW for this device. When folded, the area of the device is about 67 cm$^2$, which means that if a regular-sized window (0.5 m$^2$) is filled with these thermoelectric modules, power of 6.6 µW is produced with the 10 K temperature difference. This is enough to power the sensor node used here as an example. However, as the power production varies a lot over time, an energy storage and an algorithm to control the power consumption of the electronics are needed. As is characteristic for TEGs, raising the temperature difference increases the power very rapidly. Therefore, environments with more extreme conditions may provide interesting use cases for this solution.

The antennas, that is those for UHF RFID transponders and for Bluetooth radio, were also successfully demonstrated. UHF RFID transponder antennas implemented on a flexible Kapton NH substrate produced correct frequency response, but their sensitivity remained 2.3 dB below what was predicted by the simulations, the corresponding measured radiation efficiency being about -9.1 dB at 900 MHz. The Bluetooth antennas implemented on glass appeared to produce higher radiation efficiencies than the one with the flexible substrate. When compared to the simulation results, the ones made on 3 mm glass gave actually better efficiency values than what predicted by the simulations. The radiation efficiency values of the Bluetooth antennas varied between -3.8 dB and -0.4 dB, depending on the substrate. Higher radiation efficiency and better correspondence with the simulations of the glass antennas may be due to the AZO coating being more stable on a rigid substrate.

## Acknowledgment

The authors would like to thank R. Ritasalo from Picosun for providing the AZO coating by ALD process for the prototypes. The authors would also like to thank their colleagues at VTT: M. Hillukkala, M. Korkalainen, I. Marttila and T. Pernu for the development of the electronics of the sensor node, and R. Grenman and M. Vilkman for screen printing the silver conductors on the TEG foils. DuPont is acknowledged for bringing the experimental CS Series of Kapton available.

## References

[1] C. Buratti, A. Conti, D. Dardari, and R. Verdone, "An overview on wireless sensor networks technology and evolution," Sensors (Basel), 2009, vol. 9, no. 9, pp. 6869–96

[2] K. Z. Panatik, K. Kamardin, S. A. Shariff, S. S. Yuhaniz, N. A. Ahmad, O. M. Yusop and SA Ismail, "Energy harvesting in wireless sensor networks: A survey", *2016 IEEE 3rd International Symposium on Telecommunication Technologies (ISTT)*, 2016, pp. 53 - 58

[3] A. M. Abdal-Kadhim and K. S. Leong, "Application of thermal energy harvesting from low-level heat sources in powering up WSN node", *2017 2nd International Conference on Frontiers of Sensors Technologies (ICFST)*, 2017, pp. 131 - 135

[4] L. Hou and S. Tan, "A preliminary study of thermal energy harvesting for industrial wireless sensor networks", *2016 10th International Conference on Sensing Technology (ICST)*, 2016, pp. 1 - 5

[5] Q. L. Li, S. W. Cheung and Di Wu and T. I. Yuk, "Optically Transparent Dual-Band MIMO Antenna Using Micro-Metal Mesh Conductive Film for WLAN System", *IEEE Antennas and Wireless Propagation Letters*, 2017, Vol. 16, pp. 920 - 923

[6] M. Ruoho, T. Juntunen, T. Alasaarela, M. Pudas, and I. Tittonen, "Transparent, Flexible, and Passive Thermal Touch Panel", *Adv. Mater. Technol.* (2016) DOI: 10.1002/admt.201600204

[7] R. Correia, N. B. Carvalho, "Design of high order modulation backscatter wireless sensor for passive IoT solutions", *2016 IEEE Wireless Power Transfer Conference (WPTC)*, 2016

[8] Datasheet of Nordic Semiconductors nRF51422 Multiprotocol ANT™/Bluetooth® low energy System on Chip, accessed on May 8, 2018. [Online]. Available: http://infocenter.nordicsemi.com/pdf/nRF51422_PS_v2.1.pdf

[9] K. Tappura, "A numerical study on the design trade-offs of a thin-film thermoelectric generator for large-area applications". *Renewable Energy* vol. 120, pp. 78–87, May 2018. DOI: 10.1016/j.renene.2017.12.063

[10] K. Ellmer, "Past achievements and future challenges in the development of optically transparent electrodes", *Nature Photonics*, Vol. 6, Dec. 2012, DOI: DOI: 10.1038/NPHOTON.2012.282

[11] T. Dhakal, A. S. Nandur, R. Christian, P. Vasekar, S. Desu C. Westgate, D.I. Koukis, D.J. Aren, D.B. Tanner, "Transmittance from visible to mid infra-red in AZO films grownby atomic layer deposition system". *Solar Energy* vol. 86, 1306–1312, 2018. DOI:10.1016/j.solener.2012.01.022

[12] J. Loureiro, N. Neves, R. Barros, T. Mateus, R. Santos, S. Filonovich, et al., "Transparent aluminium zinc oxide thin films with enhanced thermoelectric properties, *J. Mater. Chem.* vol. 2, 2014, pp. 664- 6655. DOI: 10.1039/c3ta15052f

[13] T.Q. Trinh., T.T. Nguyen, D.V. Vu, D.H. Le, Structural and thermoelectric properties of Al-doped ZnO thin films grown by chemical and physical methods, *J Mater Sci: Mater Electron* (2017) 28:236–240, DOI 10.1007/s10854-016-5516-z

[14] M. E. Zamudio, T. Busani, Y. Tawk, J. Costantine and C. Christodoulou, "Design of AZO film for optically transparent antennas", *2016 IEEE International Symposium on Antennas and Propagation (APSURSI)*, 2016, pp. 127 - 128

[15] M. D. Poliks, Y. Sung, J. Lombardi, R. Malay, J. Dederick, C. R. Westgate; M. Huang, S. Garner, S. Pollard and C. Daly, "Transparent Antennas for Wireless Systems Based on Patterned Indium Tin Oxide and Flexible Glass", *2017 IEEE 67th Electronic Components and Technology Conference (ECTC)*, 2017, pp. 1443 - 1448

[16] M. Awalludin, M. T. Ali, M. H. Mamat, "Transparent antenna using aluminum doped zinc oxide for wireless application", *2015 IEEE Symposium on Computer Applications & Industrial Electronics (ISCAIE)*,

[17] T. Tynell, R. Okazaki, I. Terasaki, H. Yamauchi and M. Karppinen, "Electron doping of ALD-grown ZnO thin films through Al and P substitutions", *J. Mater. Sci.* vol. 48, 2013, pp. 2806–2811. DOI: 10.1007/s10853-012-6942-9

[18] *Datasheet of Kapton® polymimide film by DuPont™*, accessed on May 8, 2018. [Online]. http://www.dupont.com/content/dam/dupont/products-and-services/membranes-and-films/polyimde-films/documents/DEC-Kapton-general-specs.pdf and http://www.shagal-thermal.solutions/itemfiles/168940_kapton%20shagal.pdf

[19] C.-Y. Peng, M. M. Hamasha, D. VanHart, S. Lu, and C. R. Westgate, "Electrical and Optical Degradation Studies on AZO Thin Films Under Cyclic Bending Conditions", *IEEE Transactions on Device and Materials Reliability,* vol. 13, no. 1, pp. 236–244, Mar 2013.

[20] Heat Transfer and AC/DC Module User's Guides. COMSOL Multiphysics® v. 5.3. COMSOL AB, Stockholm, Sweden.

[21] K. Arapov, K. Jaakkola, V. Ermolov, G. Bex, E. Rubingh, S. Haque, H. Sandberg, R. Abbel, G. de With and H. Friedrich, "Graphene Screen-printed Radio-frequency Identification Devices on Flexible Substrates", *Physica Status Solidi – Rapid Research Letters*, Volume 10, Issue 11; Nov 2016, pp. 812 – 818

[22] P. Nikitin, K. V. S. Rao, and S. Lam, "Antenna design for UHF RFID tags: A review and a practical application," *IEEE Trans. Antennas Propag.*, vol. 53, no. 12, Dec. 2005, pp. 3870–3876

[23] *Ansys HFSS High Frequency Electromagnetic Field Simulation*, accessed on May 8, 2018. [Online]. Available: http://www.ansys.com/Products/Electronics/ANSYS-HFSS

[24] J. H. Choi, K. Ryu, K.Park, S-J Moon, "Thermal conductivity estimation of inkjet-printed silver nanoparticle ink during continuous wave laser sintering", *International Journal of Heat and Mass Transfer* vol. 85, 904–909, 2015. DOI: 10.1016/j.ijheatmasstransfer.2015.01.056.

[25] Y. Q. Cao, X. B. Zhao, T. J. Zhu, X. B. Zhang, and J. P. Tu, "Syntheses and thermoelectric properties of $Bi_2Te_3/Sb_2Te_3$ bulk nanocomposites with laminated nanostructure, *Applied Physics Letters* vol. 92, 143106, 2008. DOI: 10.1063/1.2900960.

[26] *ERC Recommendation 70–03 Relating to the Use of Short Range Devices (SRD)*, p. 34, accessed on May 8, 2018. [Online]. Available: https://www.ecodocdb.dk/download/25c41779-cd6e/Rec7003e.pdf

[27] *Datasheet of Impinj Monza R6 UHF RFID Microchip*, accessed on May 8, 2018. [Online]. Available: https://support.impinj.com/hc/article_attachments/115001963950/Monza%20R6%20Tag%20Chip%20Datasheet%20R5%2020170901.pdf

[28] *Tagformance by Voyantic*, accessed on May 8, 2018. [Online]. Available: http://voyantic.com/tagformance

[29] P. Nikitin, K. V. S. Rao, S. Lam. *UHF RFID Tag Characterization: Overview and State-of-the-Art*, accessed on May 8, 2018. [Online]. Available: https://pdfs.semanticscholar.org/4c92/ad48e34cf7ef6a11e7652819cb6cb293d9e2.pdf

[30] *Datasheet of HUBER+SUHNER 1324.19.0002 antenna*, accessed on May 8, 2018. [Online]. Available: http://www.firstsourceinc.com/DataSheets/HuberSuhner/22649580.pdf

[31] *Datasheet of ST Microelectronics balun BAL-NRF02D3,* accessed on May 8, 2018. [Online]. Available: http://www.st.com/content/ccc/resource/technical/document/datasheet/8a/a7/b0/f7/24/9f/44/9f/DM00087690.pdf/files/DM00087690.pdf/jcr:content/translations/en.DM00087690.pdf

[32] *Application Note of Fractus Bluetooth, 802.11b/g WLAN Chip Antenna by Texas Instruments*, accessed on May 8, 2018. [Online]. Available: http://www.ti.com/lit/an/swra092b/swra092b.pdf

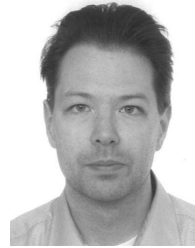

**Kaarle Jaakkola** received the Master of Science (Tech.) degree in electrical engineering from the Helsinki University of Technology (currently Aalto University), Espoo, Finland, in 2003.

Since 2000 he has been working at the VTT Technical Research Centre of Finland, currently as a Senior Scientist. His research interests and expertise include RFID systems, electronics, wireless and applied sensors, antennas, electromagnetic modelling and RF electronics. He has e.g. developed RF parts for RFID systems and designed antennas for both scientific use and commercial products. Antennas designed by him can be found in several commercial RFID transponders.

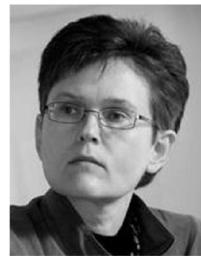

**Kirsi Tappura** received the M.Sc. (Tech.) with distinction, Lic.Sc. (Tech.), and D.Sc. (Tech.) degrees in technical physics from the Tampere University of Technology (TUT), Tampere, Finland, in 1990, 1992, and 1993, respectively.

She continued her research on semiconductor physics and optoelectronics at TUT as a Research Scientist, Project Manager and a Postdoctoral Research Fellow of the Academy of Finland until joined the Nokia Research Center as a Senior Research Scientist involved with novel electronic displays. Since late 1997, Dr. Tappura has been with the VTT Technical Research Centre of Finland, Espoo/Tampere, Finland, since 1999 as a Senior Scientist and, since 2011, as a Principal Scientist serving also as a Team Leader of modelling, sensors and energy materials related teams during 1999-2001 and 2006-2012. Since 1999, she has also been a Docent of Physics with TUT. Dr. Tappura is currently a Principal Scientist with VTT. Her research interests include the optical (including plasmonic), electronic and thermal properties of various sensing, detector/imaging and energy harvesting devices with an emphasis on computational physics.